\documentclass[aip,jcp,showpacs,showkeys,preprint,floatfix,superscriptaddress]{revtex4-1}

\usepackage{amssymb}
\usepackage{amsmath}
\usepackage{graphicx}
\usepackage{amsbsy}
\usepackage{epic,eepic}

\newcommand{\bq}{\begin{eqnarray}}
\newcommand{\eq}{\end{eqnarray}}
\newcommand{\bqn}{\begin{eqnarray*}}
\newcommand{\eqn}{\end{eqnarray*}}
\newcommand{\rr}{{\mathbf r}}

\begin{document}
\title{Andersen-Weeks-Chandler perturbation theory and
  one-component sticky-hard-spheres} 

\author{Riccardo Fantoni}
\email{rfantoni@ts.infn.it}
\affiliation{Universit\`a di Trieste, Dipartimento di Fisica, strada
  Costiera 11, 34151 Grignano (Trieste), Italy} 
\date{\today}

\begin{abstract}
We apply second order Andersen-Weeks-Chandler perturbation
theory to the one-component sticky-hard-spheres fluid. We
compare the results with the mean spherical approximation, the
Percus-Yevick approximation, two generalized Percus-Yevick
approximations, and the Monte Carlo simulations.   
\end{abstract}


\maketitle
\section{Introduction}
\label{sec:intro}

The sticky-hard-sphere (SHS) model introduced by R. J. Baxter in 1968 
\cite{Baxter68} plays an important role in soft matter offering a
description of a sterically stabilized colloidal suspension
\cite{Fantoni05b,Fantoni06a,Fantoni06b,Fantoni06c,Fantoni07,Fantoni08a,Fantoni09b}. 

In this work we apply Andersen-Weeks-Chandler (AWC)
thermodynamic-perturbation-theory (TPT) \cite{Hansen} to treat the SHS 
three-dimensional fluid and we compare the results for the equation of
state of our calculation with the ones for the
mean-spherical-approximation (MSA) \cite{Hansen}, for the
Percus-Yevick (PY) approximation \cite{Hansen}, for two 
generalized-Percus-Yevick (GPY) approximations (C0 and C1 in 
Ref. \cite{Gazzillo04}), and for the Monte Carlo simulations of Miller
and Frenkel \cite{Miller03}.

We are then able to show how the TPT breaks down at low reduced
temperature and high density. Our analysis gives a reference
benchmark for the behavior of the SHS system when treated with the AWC
TPT scheme.

Our analysis also clarifies the role played by the reducible Mayer
diagrams in the second order AWC TPT.
 
The work is organized as follows. In Section \ref{sec:wca} we
introduce the AWC TPT scheme, in Section \ref{sec:shs} we define the
SHS fluid model, in Sections \ref{sec:calculation} we outline our
calculation of the AWC TPT for the SHS fluid, in Section
\ref{sec:reducible} we clarify the role played by the reducible
integrals, in Section \ref{sec:technical} we discuss some technical
details regarding our Monte Carlo calculation of the various order
terms of the TPT, in Section \ref{sec:results} we present our results,
and Section \ref{sec:discussion} is for our conclusive discussion.

\section{The Andersen-Weeks-Chandler thermodynamic perturbation scheme}
\label{sec:wca}
Following AWC perturbation theory \cite{Andersen71} we consider the 
Helmholtz free energy $A$ as a functional of the Boltzmann factor
$e(1,2)=\exp[-\beta\phi(1,2)]$ ($\phi(1,2)$ being the pair
interaction potential of the fluid under exam) and expand it in a
Taylor series around the Boltzmann factor, $e_0(1,2)$, of a given
reference system. Working in the grand-canonical ensemble we obtain
the following perturbative expansion in $\Delta e=e-e_0$
\bq \label{wca}
\beta (A[e]-A[e_0])&=&\beta[\Delta A]_{(1)}+\beta[\Delta
A]_{(2)}+\ldots~,\\
\label{wca1}
\beta[\Delta A]_{(1)}&=&-\frac{1}{2}\int d1d2\,\frac{\rho_0(1,2)}{e_0(1,2)}\,
\Delta e(1,2)~,\\ \nonumber
\beta[\Delta A]_{(2)}&=&-\frac{1}{2}\left[\int d1d2d3\,
\frac{\rho_0(1,2,3)}{e_0(1,2)e_0(1,3)}\,\Delta e(1,2)\Delta
e(1,3)+\right.\\ \nonumber
&&\left.\frac{1}{4}\int d1d2d3d4\,\frac{\rho_0(1,2,3,4)-\rho_0(1,2)\rho_0(3,4)}
{e_0(1,2)e_0(3,4)}\,\Delta e(1,2)\Delta e(3,4)\right]+\\ \label{wca2}
&&\frac{1}{2\bar{N}}\left(\rho^2\frac{\chi_T^0}{\chi_T^{id}}\right)\left\{
\frac{\partial}{\partial\rho}\beta[\Delta A]_{(1)}\right\}^2~.
\eq
where $\beta=1/(k_BT)$ (with $k_B$ Boltzmann constant and $T$ absolute
temperature), $\bar{N}$ average number of particles, $\rho=\bar{N}/V$
(with $V$ volume of the system), $\chi_T^{id}=\beta/\rho$ isothermal
compressibility of the ideal gas, $\chi_T^0$ isothermal
compressibility of the reference system, $\rho_0(1,\ldots,n)$ the
grand-canonical ensemble $n-$body correlation function of the
reference system, and in the last term of Eq. (\ref{wca2}) the density
derivative is taken at constant temperature, volume, and chemical
potential. In order to derive these expressions one can adapt the
details found in Appendix D of Hansen and McDonald book \cite{Hansen}
where their expression (6.2.14) is found. It is then an easy task to
pass from their expansion in terms of the pair-potential variation to
our expansion in terms of the Boltzmann factor variation.

\section{One-component sticky-hard-spheres}
\label{sec:shs}
For the Baxter \cite{Baxter68} one-component sticky-hard-spheres (SHS)
model one has 
\bq
e(r)=\theta(r-\sigma)+\frac{\sigma}{12\tau}\delta(r-\sigma)~,
\eq 
where $\sigma$ is the spheres diameter, $\tau$ the reduced
temperature, $\theta$ is the Heaviside step function, and $\delta$ the
Dirac delta function. 

Choosing as reference system the hard-spheres (HS) model one has
\bq
e_0(r)=\theta(r-\sigma)~,
\eq
so that 
\bq
\Delta e(r)=\frac{\sigma}{12\tau}\delta(r-\sigma)~.
\eq
So one sees that AWC expansion (\ref{wca}) reduces to an expansion in
powers of $1/\tau$.

\section{Calculation}
\label{sec:calculation}
Before expression (\ref{wca2}) can be used some approximation must be
introduced for the three- and four-body distribution functions. The
most widely used approximation is Kirkwood superposition approximation
\cite{Kirkwood35}. This has previously successfully applied to the
second order thermodynamic perturbation study of the square well
potential by Henderson and Barker \cite{Henderson71}. 

Using the Kirkwood superposition approximation (KSA) \cite{Kirkwood35}
one can express the $n-$body correlation functions
$\rho_0(1,\ldots,n)=\rho^ng_0(1,\ldots,n)$ in terms of pair
distribution functions according to
\bq \label{super}
g_0(1,\ldots,n)\approx\prod_{i<j}^ng_0(i,j)~.
\eq
The idea is to use for the pair distribution function of the reference
HS system the analytic solution of the Ornstein-Zernike equation with
the Percus-Yevick closure.

The first two terms in the perturbative expansion (\ref{wca}) reduce
to 
\bq
\beta\frac{[\Delta A]_{(1)}}{\bar{N}}&=&-\frac{I_2}{\rho}~,\\ \label{wca2a}
\beta\frac{[\Delta A]_{(2)}}{\bar{N}}&=&-\frac{1}{2}
\left(\frac{I_3}{\rho}+\frac{I_4}{\rho}\right)+
\frac{1}{2}\left(\frac{\chi_T^0}{\chi_T^{id}}\right)
\left(\frac{\partial I_2}{\partial\rho}\right)^2~,
\eq
where
\bq
\frac{I_2}{\rho}=\frac{1}{2\rho}\frac{1}{V}\int d1
d2\,\frac{\rho_0(1,2)}{e_0(1,2)}\,\Delta e(1,2)=
\frac{1}{\tau}(\eta\bar{y}_0)~,
\eq
where $\eta=\frac{\pi}{6}\rho\sigma^3$ is the hard sphere packing
fraction, $y_0(1,2)=g_0(1,2)/e_0(1,2)$ is the cavity function of the
reference system and $\bar{y}_0=y_0(|\rr_1-\rr_2|=\sigma)$. 
Upon using KSA one finds,
\bq
\frac{I_3}{\rho}&=&\frac{1}{\rho}\frac{1}{V}\int d1d2d3\,
\frac{\rho_0(1,2,3)}{e_0(1,2)e_0(1,3)}\,\Delta e(1,2)\Delta
e(1,3)\\ \nonumber
&\approx&\frac{\rho^2}{V}\int d1d2d3\,y_0(1,2)y_0(1,3)J_3(1,2,3)
\Delta e(1,2)\Delta e(1,3)~,\\
\frac{I_4}{\rho}&=&\frac{1}{4\rho}\frac{1}{V}\int d1d2d3d4\,
\frac{\rho_0(1,2,3,4)-\rho_0(1,2)\rho_0(3,4)}
{e_0(1,2)e_0(3,4)}\,\Delta e(1,2)\Delta e(3,4)\\ \nonumber
&\approx&\frac{\rho^3}{4V}\int d1d2d3d4\,y_0(1,2)y_0(3,4)J_4(1,2,3,4)
\Delta e(1,2)\Delta e(3,4)~,
\eq
where we have introduced 
\bq
J_3(1,2,3)&=&1+h_0(2,3)~,\\ \nonumber
J_4(1,2,3,4)&=&4h_0(1,3)+ \\ \nonumber
&&4h_0(1,3)h_0(1,4)+ \\ \nonumber
&&2h_0(1,4)h_0(2,3)+ \\ \nonumber
&&4h_0(1,3)h_0(1,4)h_0(2,3)+ \\
&&h_0(1,3)h_0(1,4)h_0(2,3)h_0(2,4)~,
\eq
where $h_0(1,2)=g_0(1,2)-1$ is the total correlation function of the
reference system. Note that the first term in $J_3$ and the first and
second terms in $J_4$ give rise to {\sl reducible} integrals
(i.e. integrals that can be reduced into products of simpler
integrals).

It is convenient to perform the calculation of $I_3$ and $I_4$ in
reciprocal space, to get,
\bq
\frac{I_3}{\rho}&\approx&\frac{1}{\tau^2}(2\eta\bar{y}_0)^2
\left(1+\frac{1}{12\pi}\frac{1}{\eta}g_1\right)~,\\
\frac{I_4}{\rho}&\approx&\frac{1}{\tau^2}(2\eta\bar{y}_0)^2
\frac{1}{4}\left[4\left(\frac{1}{a^2}-1\right)+
\frac{1}{3\pi}\frac{1}{\eta}h_2^a+
\frac{1}{6\pi}\frac{1}{\eta}h_2^b+
\frac{1}{72\pi^2}\frac{1}{\eta^2}h_3+
\frac{1}{6^32^6\pi^4}\frac{1}{\eta^3}h_4\right]~,
\eq
and 
\bq \label{g1}
g_1&=&\int_0^\infty dz\,z^2j_0^2(z)H(z)~,\\
h_2^a&=&\int_0^\infty dz\,z^2j_0(z)H^2(z)~,\\
h_2^b&=&\int_0^\infty dz\,z^2j_0^2(z)H^2(z)~,\\
h_3&=&\int_0^\infty dz_1\,z_1^2\int_0^\infty dz_2\,z_2^2\int_{-1}^1dx\, 
j_0(z_1)j_0(z_2)H(z_1)H(z_2)H(\sqrt{z_1^2+z_2^2-2z_1z_2x})~,\\ \nonumber
h_4&=&\int_0^\infty dz_1\,z_1^2\int_0^\infty dz_2\,z_2^2
\int_0^\infty dz_3\,z_3^2\int_0^\pi d\theta_1\,\sin\theta_1
\int_0^\pi d\theta_2\,\sin\theta_2\int_0^{2\pi} d\phi\,\\ \nonumber
&&j_0(z_1)j_0(\sqrt{z_2^2+z_3^2-2z_2z_3\cos\theta_2})
H(z_2)H(z_3)H(\sqrt{z_1^2+z_2^2-2z_1z_2\cos\delta})\\ \label{h4}
&&H(\sqrt{z_1^2+z_3^2-2z_1z_3\cos\theta_1})~,
\eq
where in the integrand of $h_4$ 
\bq
\cos\delta=\cos\theta_1\cos\theta_2+\sin\theta_1\sin\theta_2\cos\phi~.
\eq

In all these expressions we have introduced the following notation
\bq
a^2&=&\frac{\chi_T^{id}}{\chi_T^0}=1-\rho\tilde{c}_0(0)~,\\
\bar{y}_0&=&y_0(\sigma)=g_0(\sigma)/e_0(\sigma)~,\\
H(z)&=&\rho\tilde{h}_0(z/\sigma)=
\frac{\rho\tilde{c}_0(z/\sigma)}{1-\rho\tilde{c}_0(z/\sigma)}~,\\
j_0(z)&=&\frac{\sin z}{z}~,
\eq
where $g_0(r), y_0(r),\tilde{h}_0(k), \tilde{c}_0(k)$ are respectively
the hard spheres 
radial distribution function, cavity function, the Fourier transform
of the total correlation function and the Fourier transform of the
direct correlation function, and $j_0$ is the zeroth order spherical
Bessel function of the first kind.

Finally the Fourier transform of the HS direct correlation function
calculated through the Percus-Yevick closure is given by
\cite{Ashcroft66} 
\bq
\rho\tilde{c}_0(z/\sigma)\approx-24\eta\int_0^1ds\,s^2j_0(sz)
(\alpha+\beta s+\gamma s^3)~,
\eq 
where
\bq
\alpha&=&\left[\frac{1+2\eta}{(1-\eta)^2}\right]^2~,\\
\beta &=&-6\eta\left[\frac{1+\eta/2}{(1-\eta)^2}\right]^2~,\\
\gamma&=&\frac{\eta}{2}\left[\frac{1+2\eta}{(1-\eta)^2}\right]^2~.
\eq
and it is easily verified that under such approximation one has
\bq
a&\approx&\frac{1+2\eta}{(1-\eta)^2}~,\\
\bar{y}_0&\approx&\frac{1+\eta/2}{(1-\eta)^2}~.
\eq

\section{Neglecting reducible integrals}
\label{sec:reducible}
It has been observed by Henderson and Barker \cite{Henderson71} that
the role of the last term in Eq. (\ref{wca2})
\bq \label{corc}
{\cal C}\bar{N}=\frac{1}{2\bar{N}}\left(\rho^2
\frac{\chi_T^0}{\chi_T^{id}}\right)\left\{\frac{\partial}{\partial\rho}
\beta[\Delta A]_{(1)}\right\}^2~,
\eq
is to cancel in the second order term of the perturbative expansion,
$[\Delta A]_{(2)}$, all reducible integrals appearing in $I_3$ and
$I_4$. So that the final expression for the second order term of 
expansion (\ref{wca}) would be (exactly the expression found in
\cite{Andersen71})
\bq \label{wca2b}
\beta\frac{[\Delta A]_{(2)}^\prime}{\bar{N}}&=&-\frac{1}{2}
\left(\frac{I_3^\prime}{\rho}+\frac{I_4^\prime}{\rho}\right)~,
\eq
where
\bq
\frac{I_3^\prime}{\rho}&=&\frac{I_3}{\rho}-\frac{1}{\tau^2}
(2\eta\bar{y}_0)^2~,\\
\frac{I_4^\prime}{\rho}&=&\frac{I_4}{\rho}-\frac{1}{\tau^2}
(2\eta\bar{y}_0)^2
\frac{1}{4}\left[4\left(\frac{1}{a^2}-1\right)+
\frac{1}{3\pi}\frac{1}{\eta}h_2^a\right]~.
\eq

Alternatively one may use the sum rule
\bq \label{gcsum3}
\frac{\partial\rho_0(1,2)}{\partial\rho}=\frac{1}{\rho}
\frac{\chi_T^{id}}{\chi_T^0}\left\{2\rho_0(1,2)+\int d3
[\rho_0(1,2,3)-\rho\rho_0(1,2)]\right\}~,
\eq
to rewrite ${\cal C}$ [Eq. (\ref{corc})] in terms of two and three
body correlation functions and upon using the superposition
approximation one finds 
\bq \label{wca2c}
\beta\frac{[\Delta A]_{(2)}^{\prime\prime}}{\bar{N}}&=&-\frac{1}{2}
\left(\frac{I_3^\prime}{\rho}+\frac{I_4^\prime}{\rho}\right)+
\frac{1}{\tau^2}\frac{a^2}{8}(2\eta\bar{y}_0)^2
\left(\frac{1}{12\pi}\frac{1}{\eta}h_2^a\right)^2~,
\eq

\section{Technical details}
\label{sec:technical}
The five integrals (\ref{g1})-(\ref{h4}) where all calculated using
Monte Carlo technique \cite{Kalos} averaging the various integrands on
$10^6$ randomly sampled points. Since all of those
integrals are improper (extending up to infinity in the $z$ variables)
it was necessary to split each integration on the $z$ variables into an
integral over $[0,1]$ plus an integral over $[1,\infty]$. This
latter integral was then reduced through a change of variable $z\to 1/z$
into an integral over $[0,1]$.

The errors on the estimate of a given integral was calculated so that
the true value of the integral would lie $99.7\%$ of the time within
the estimate plus or minus the error.

\section{Results}
\label{sec:results}
\begin{figure}[hbt]
\begin{center}
\includegraphics[width=16cm]{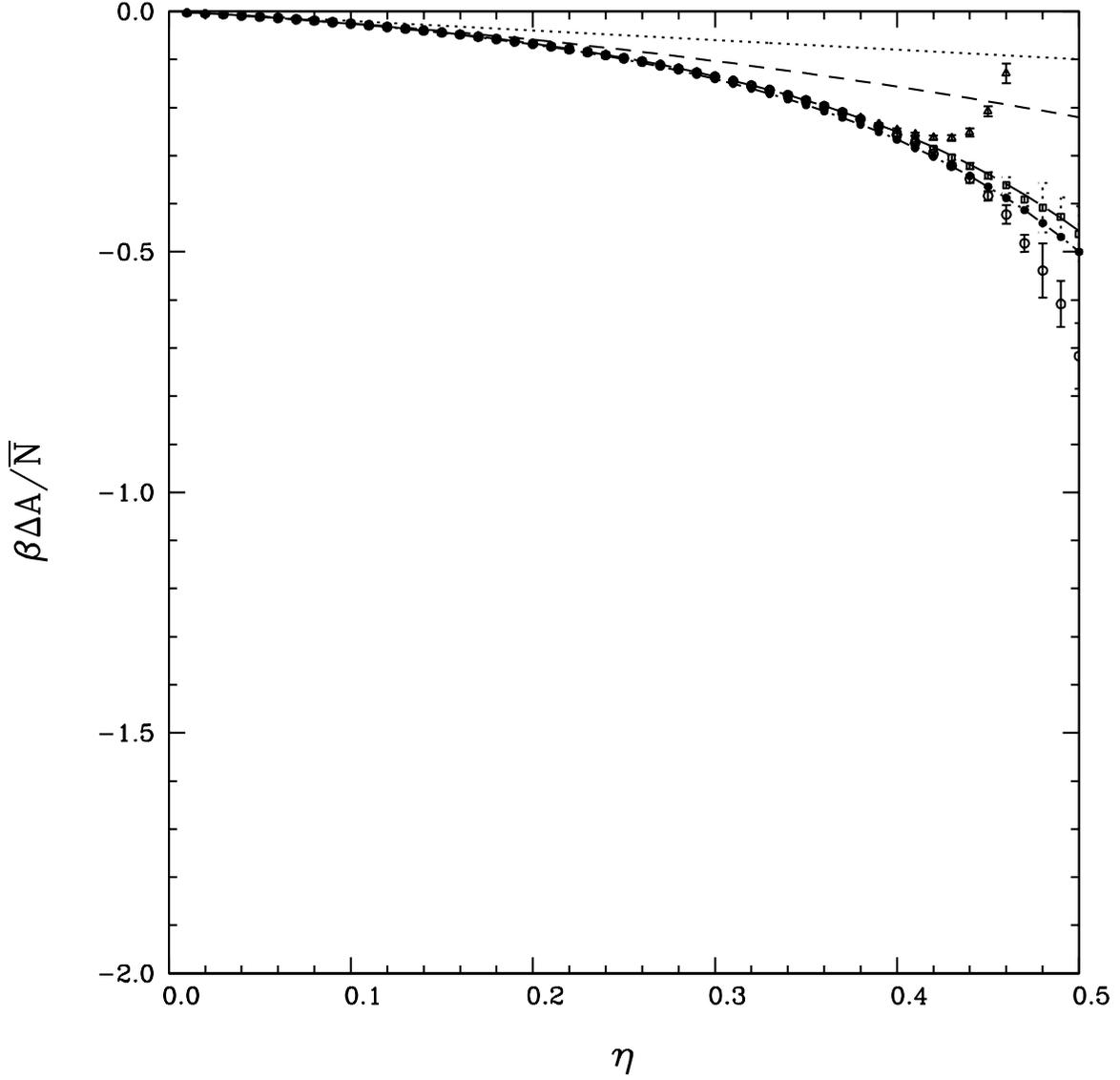}
\end{center}
\caption[]{We show $\beta\Delta
A/\bar{N}=\beta(A^{SHS}-A^{HS})/\bar{N}$ as a function of the packing
fraction at $\tau=5$ for various approximations:
(in the MSA $\Delta A=0$) C0 (dotted line) \cite{Gazzillo04}, C1
(short dashed line) \cite{Gazzillo04}, PY
(long dashed line) \cite{Hansen}, $\beta[\Delta A]_{(1)}/\bar{N}$
(dotted dashed 
line and filled circles), $\beta([\Delta A]_{(1)}+[\Delta
A]_{(2)})/\bar{N}$ (empty circles), $\beta([\Delta
A]_{(1)}+[\Delta A]_{(2)}^\prime)/\bar{N}$ (empty squares), and
$\beta([\Delta A]_{(1)}+[\Delta A]_{(2)}^{\prime\prime})/\bar{N}$
(empty triangles).   
\label{fig:wca2-5.0}
}
\end{figure}

\begin{figure}[hbt]
\begin{center}
\includegraphics[width=16cm]{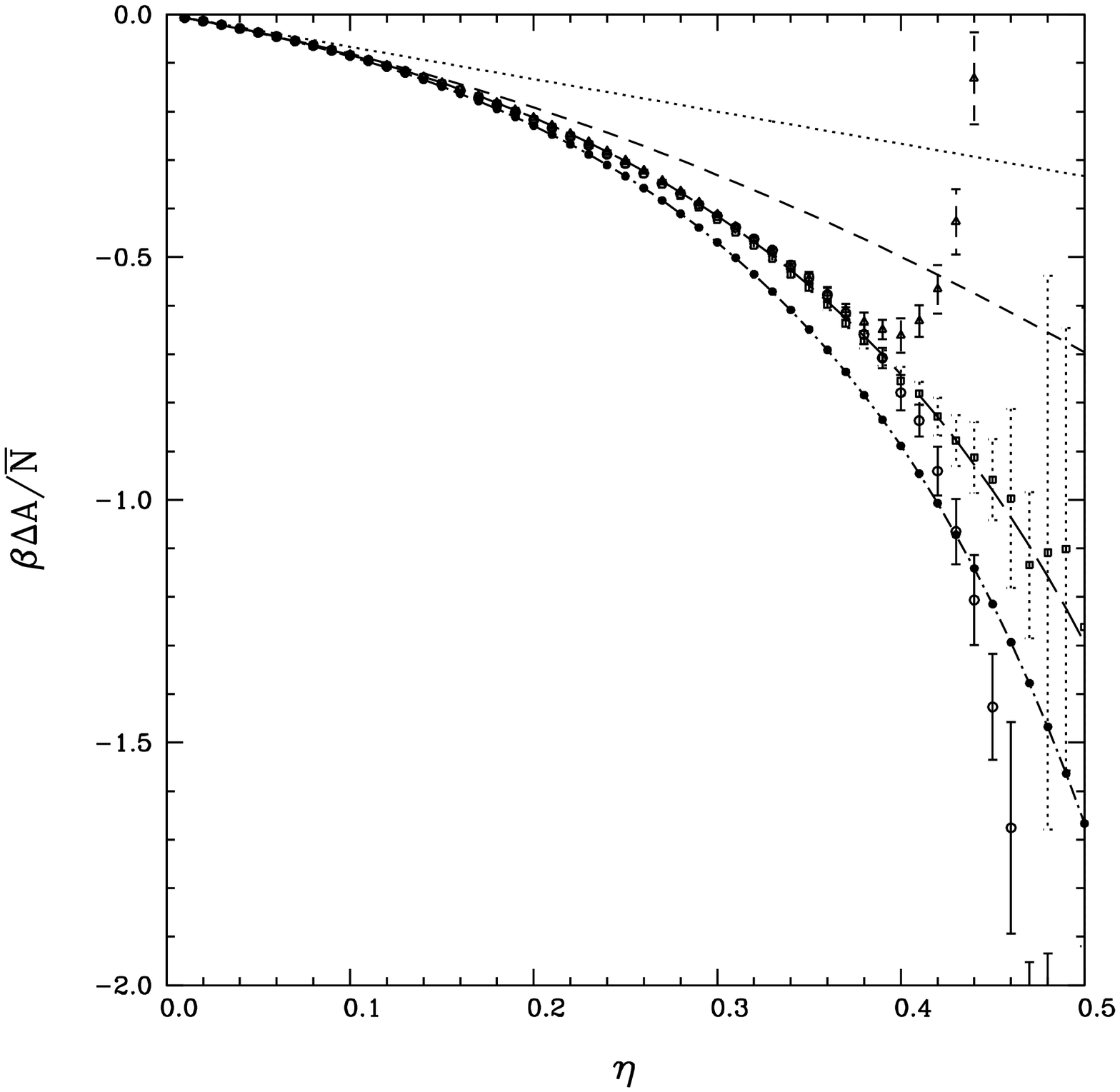}
\end{center}
\caption[]{Same as Fig. \ref{fig:wca2-5.0} at $\tau=1.5$.    
\label{fig:wca2-1.5}
}
\end{figure}

\begin{figure}[hbt]
\begin{center}
\includegraphics[width=16cm]{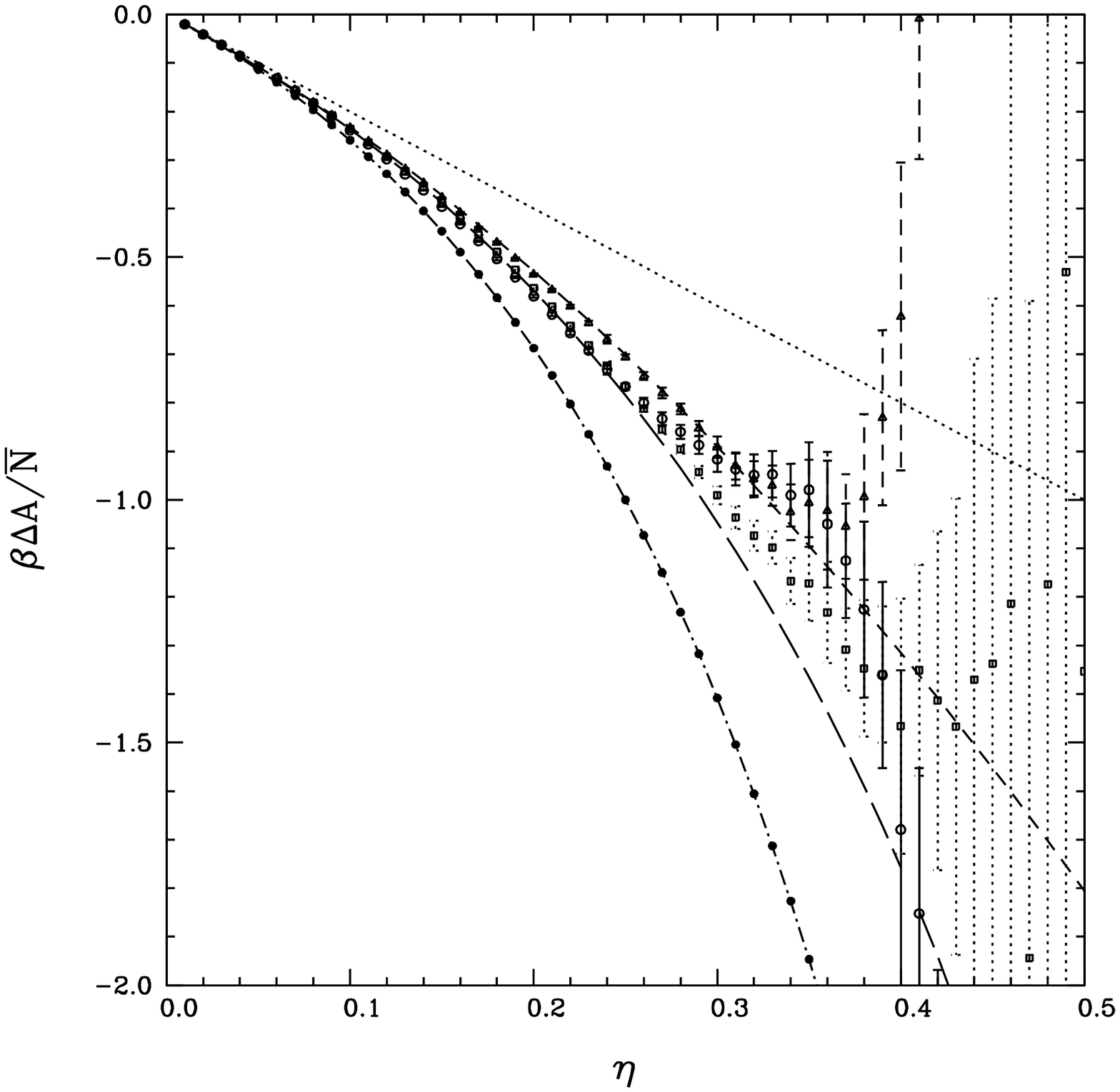}
\end{center}
\caption[]{Same as Fig. \ref{fig:wca2-5.0} at $\tau=0.5$. 
\label{fig:wca2-0.5}
}
\end{figure}

\begin{figure}[hbt]
\begin{center}
\includegraphics[width=16cm]{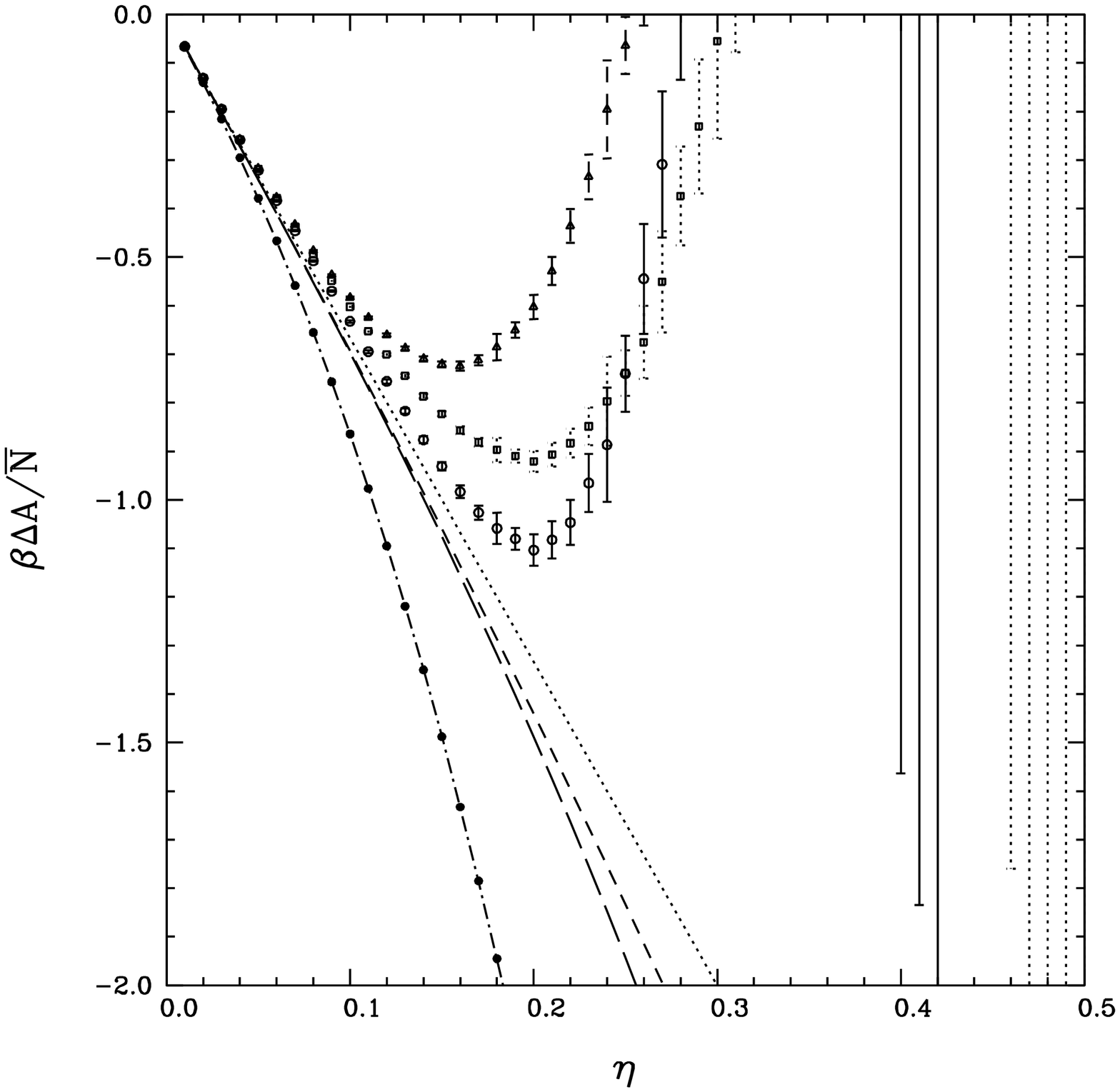}
\end{center}
\caption[]{Same as Fig. \ref{fig:wca2-5.0} at $\tau=0.15$. 
\label{fig:wca2-0.15}
}
\end{figure}

Figs. \ref{fig:wca2-5.0}-\ref{fig:wca2-0.15} show the results for
$\beta\Delta A/N$ as a function of $\eta$. Amongst
the three expressions used: (\ref{wca2a}), (\ref{wca2b}), and
(\ref{wca2c})), the more accurate is $[\Delta A]_{(2)}^\prime$, the one
suggested in \cite{Andersen71} and it falls on the PY approximation
for big $\tau$ and small $\eta$. At high $\eta$ the error bars become
more relevant. 

\begin{figure}[hbt]
\begin{center}
\includegraphics[width=16cm]{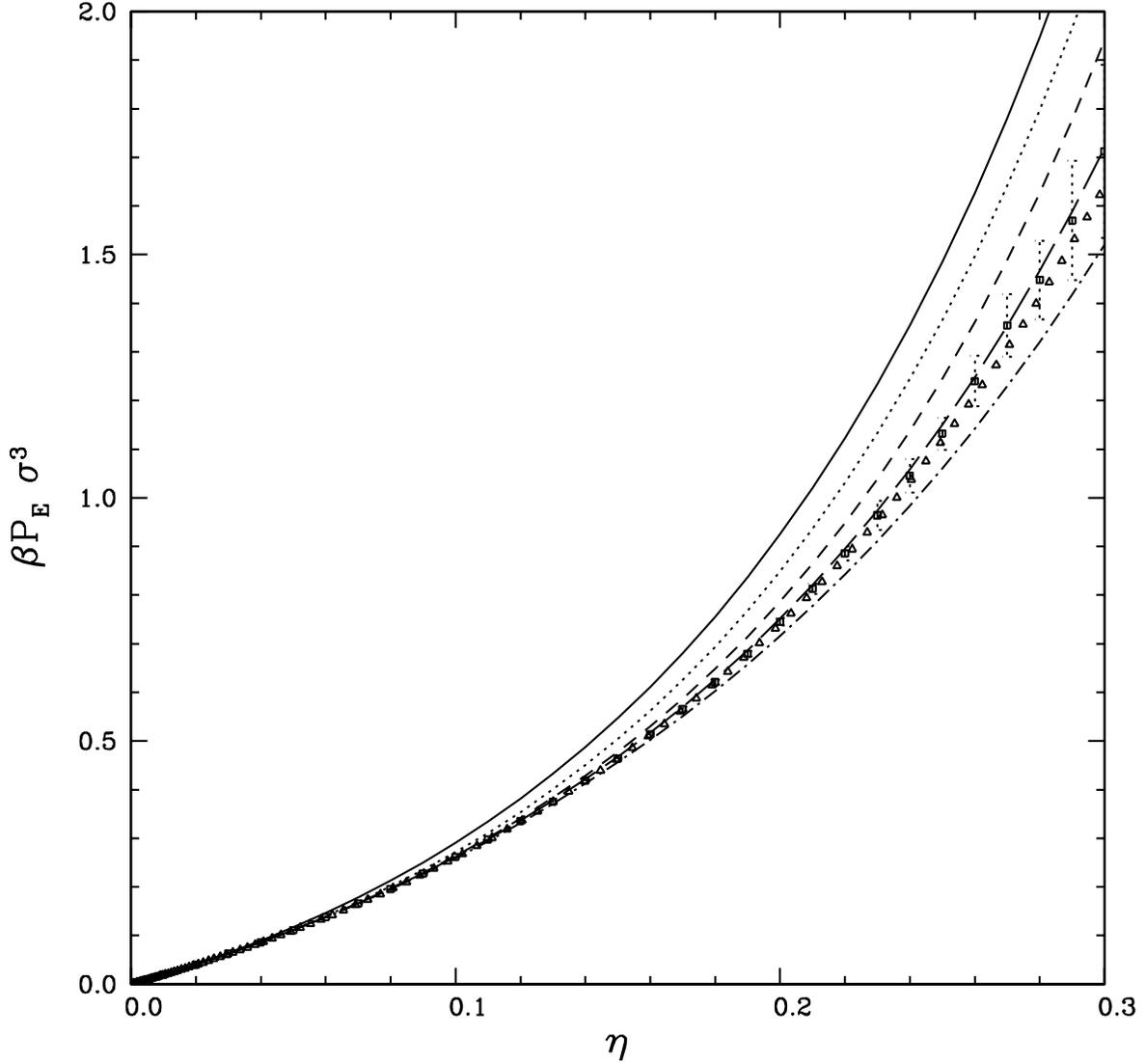}
\end{center}
\caption[]{We show $\beta P\sigma^3$ as a function of the packing
fraction at $\tau=1$ for various approximations:
MSA (continuous line), C0 (dotted line) \cite{Gazzillo04}, C1 (short
dashed line) \cite{Gazzillo04}, PY
(long dashed line) \cite{Hansen}, AWC 1st order (dotted dashed
line), AWC 2nd order (empty squares), and Monte Carlo results of
Miller and Frenkel (empty triangles) \cite{Miller03}.
\label{fig:bps3-1.0}
}
\end{figure}

\begin{figure}[hbt]
\begin{center}
\includegraphics[width=16cm]{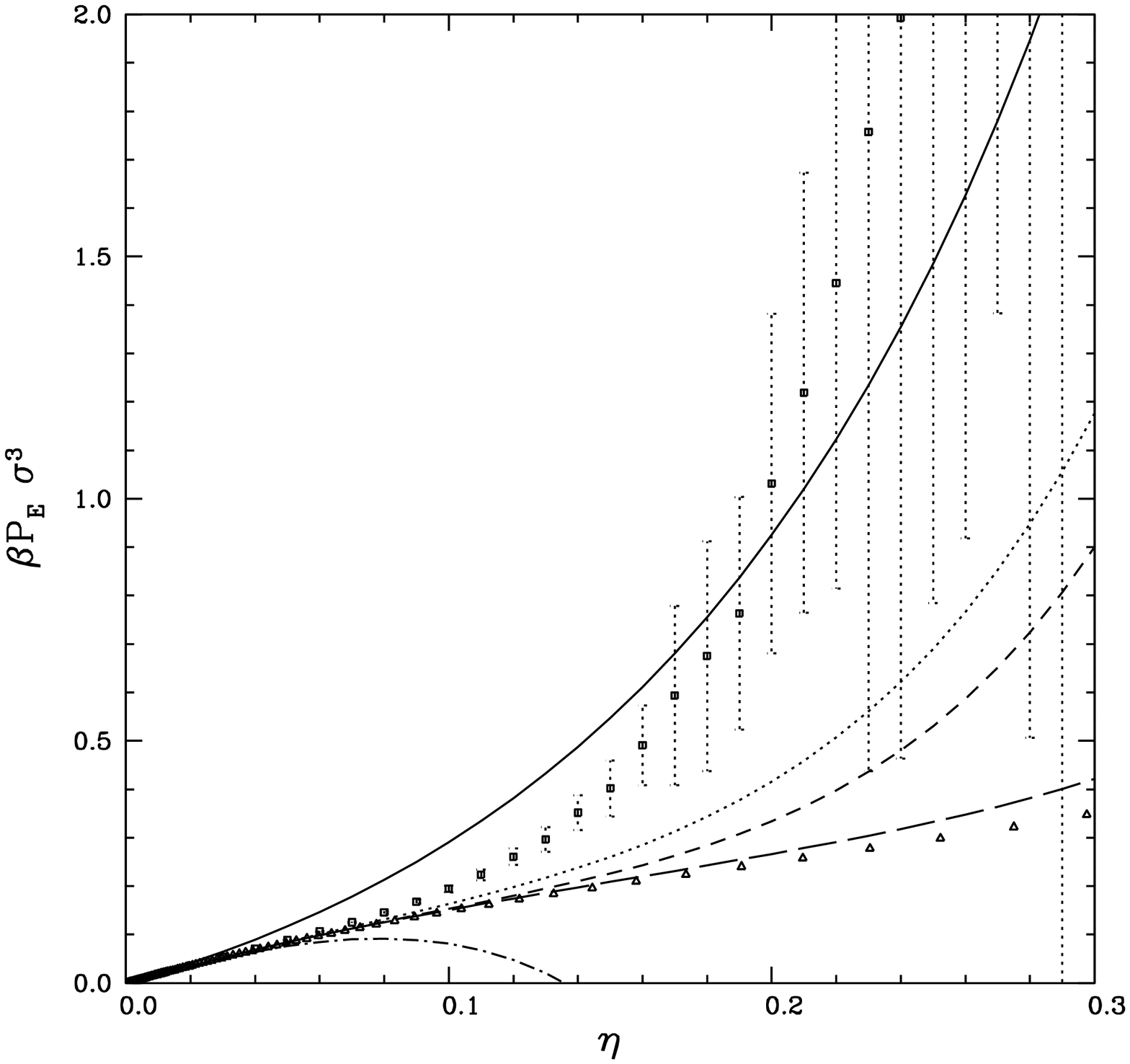}
\end{center}
\caption[]{Same as Fig. \ref{fig:bps3-1.0} at $\tau=0.15$. 
\label{fig:bps3-0.15}
}
\end{figure}

Figs. \ref{fig:bps3-0.15}-\ref{fig:bps3-1.0} show the results for
\bq
\beta P\sigma^3=\beta P_{HS}\sigma^3+\frac{6}{\pi}\eta^2
\frac{\partial\beta\Delta A/N}{\partial\eta}~,
\eq 
as a function of $\eta$, where for the pressure of the HS reference
system we chose the PY result from the compressibility route, i.e.
\bq
\beta P_{HS}\sigma^3=\frac{6}{\pi}\eta\left[
\frac{1+\eta+\eta^2}{(1-\eta)^3}\right]~.
\eq
The second order AWC TPT is taken from the (\ref{wca2b} calculation.

\begin{figure}[hbt]
\begin{center}
\includegraphics[width=16cm]{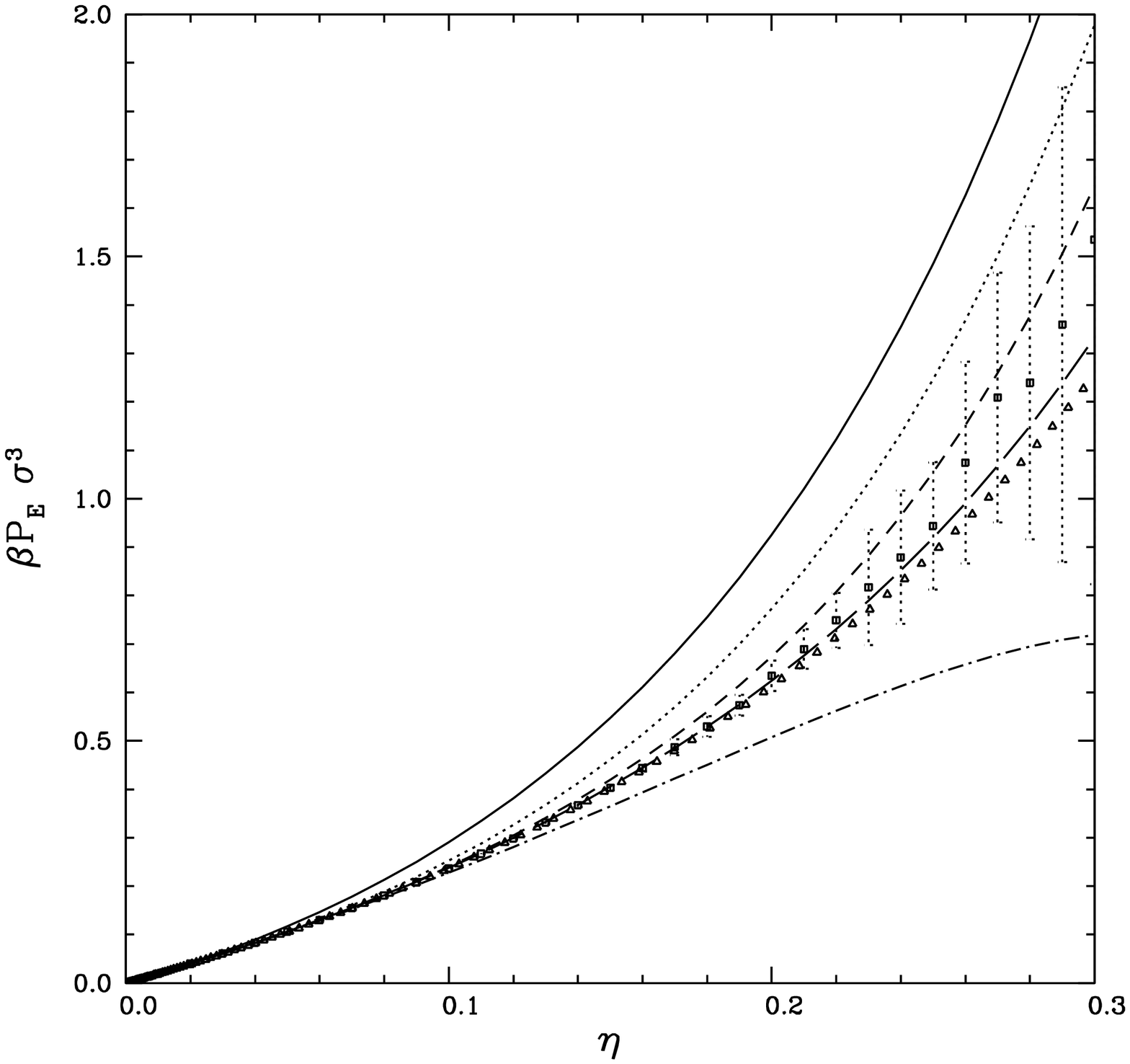}
\end{center}
\caption[]{Same as Fig. \ref{fig:bps3-1.0} at $\tau=0.5$. 
\label{fig:bps3-0.5}
}
\end{figure}
\begin{figure}[hbt]
\begin{center}
\includegraphics[width=16cm]{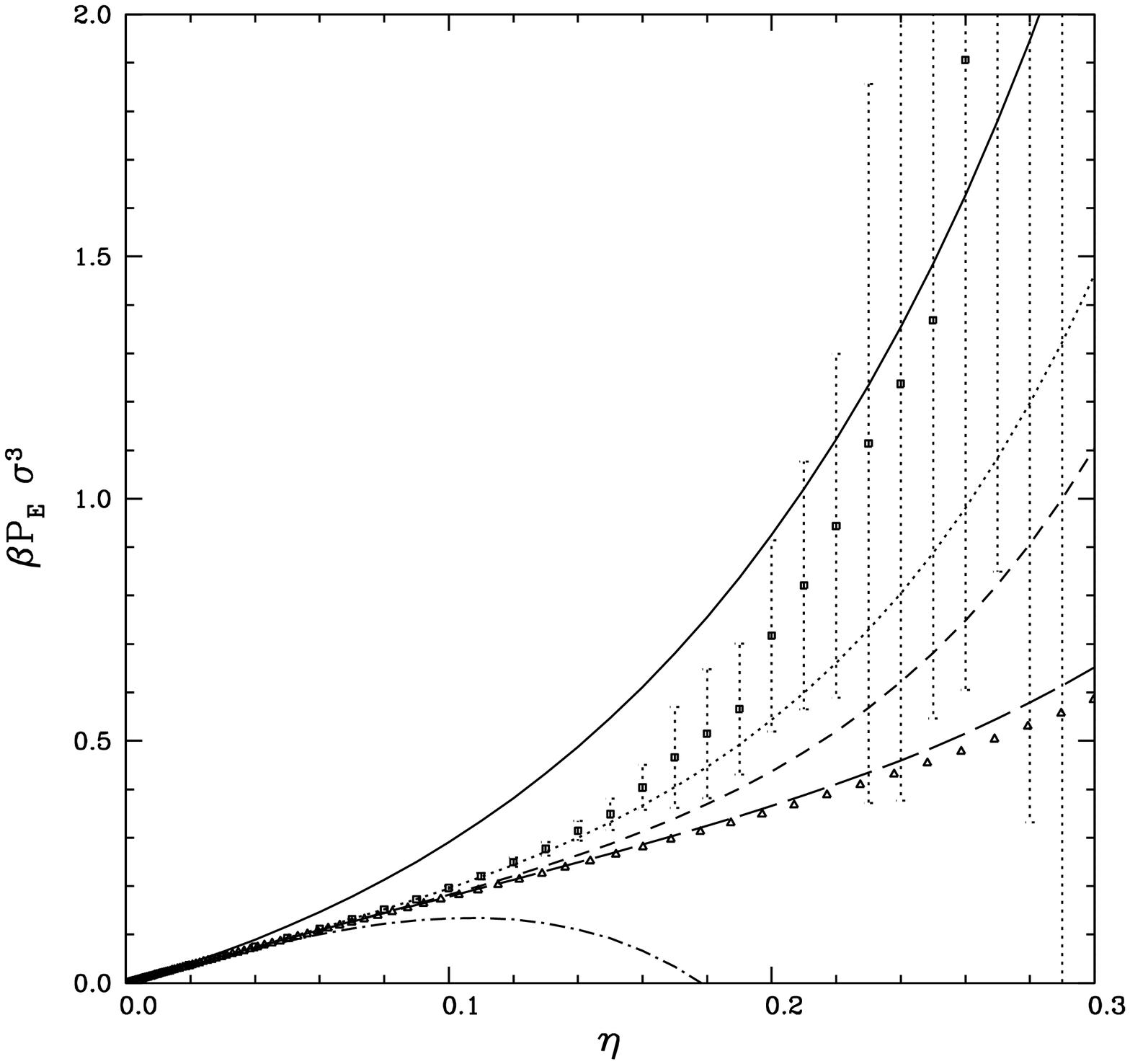}
\end{center}
\caption[]{Same as Fig. \ref{fig:bps3-1.0} at $\tau=0.2$. 
\label{fig:bps3-0.2}
}
\end{figure}

\section{Discussion}
\label{sec:discussion}
Our first calculation, the one using $[\Delta A]_{(2)}$ (see
Eq. (\ref{wca2a})) is certainly not correct because we are using the
KSA only on the integrands of the first two integrals of
Eq. (\ref{wca2}) calculating the last term exactly; this certainly
leads to an inconsistency in the use of KSA.

Our third calculation, the one using $[\Delta A]_{(2)}^{\prime\prime}$ (see
Eq. (\ref{wca2c})) is also not correct. This can be understood as
follows. It is well known that KSA fails to satisfy the sum rule
(\ref{gcsum3}). Using KSA in the left hand side of Eq. (\ref{gcsum3})
one finds  
\bq \label{gcsum3-ksa}
\frac{1}{\rho}
\frac{\chi_T^{id}}{\chi_T^0}\left\{2\rho_0(1,2)+\int d3\,
[\rho_0(1,2,3)-\rho\rho_0(1,2)]\right\}\approx
\gamma_1+\gamma_2~,
\eq
where
\bq
\gamma_1&=&g_0(1,2)2\rho ~,\\
\gamma_2&=&g_0(1,2)\frac{\chi_T^{id}}{\chi_T^0}\int d3\,
[\rho h_0(1,3)][\rho h_0(2,3)]~, 
\eq
and we used the compressibility sum rule,
\bq
\frac{\chi_T^{0}}{\chi_T^{id}}=1+\rho\int h_0(1,2)\,d1~.
\eq

Eq. (\ref{gcsum3-ksa}) can be also rewritten as,
\bq \label{dgdrksa}
\frac{\partial\ln g_0(1,2)}{\partial\rho}\approx
\frac{\chi_T^{id}}{\chi_T^0}\int d3\, h_0(1,3)h_0(2,3)~.
\eq
This approximation is certainly valid in the limit of small densities
when $\chi_T^0\to\chi_T^{id}$ and $h_0\to e_0-1=f_0$ ($f_0$ being the
Mayer function of the reference system), after all the KSA becomes
exact in such limit (as the potential of mean force tends to the pair
interaction potential). Otherwise the correction term
$\gamma_3/(\rho^2 g_0)$ would be of order $\rho$ as $\rho\to 0$ (see
the Appendix). So that the exact expression for the density derivative
of the two body correlation function would be
\bq
\frac{\partial\rho_0(1,2)}{\partial\rho}=
\gamma_1(1,2)+\gamma_2(1,2)+\gamma_3(1,2)~, 
\eq
where $\gamma_i=O(\rho^i)$ as $\rho\to 0$. It is then clear that in
calculating the square
\bq
\left[\frac{\partial}{\partial\rho}\frac{1}{2}\int d1 d2\,
\frac{\rho_0(1,2)}{e_0(1,2)}\Delta e(1,2)\right]^2~,
\eq
in the ${\cal C}$ term, the term stemming from 
\bq \label{neg}
\left[\frac{1}{2}\int d1 d2\,\frac{\gamma_2(1,2)}{e_0(1,2)}\Delta
e(1,2)\right]^2~, 
\eq
which gives rise to the last term in Eq. (\ref{wca2c}), will be of the
same leading order ($\rho^4$) as the one coming from
\bq
\left[\frac{1}{2}\int d1 d2\,\frac{\gamma_1(1,2)}{e_0(1,2)}\Delta
e(1,2)\right]
\left[\frac{1}{2}\int d1 d2\,\frac{\gamma_3(1,2)}{e_0(1,2)}\Delta
e(1,2)\right]~,
\eq
in the small density limit. But since in KSA this last term is
neglected, in order to be consistent (up to orders $\rho^3$ in the
small density limit) one needs to neglect also the term of
Eq. (\ref{neg}). Moreover it can be easily verified that the two terms
coming from $\gamma_1$ times $\gamma_1$ cancel the first reducible
integral in $I_3$ and the first reducible integral in $I_4$ whereas
the term coming from $\gamma_1$ times $\gamma_2$ cancels the second
reducible integral in $I_4$. So that Eq. (\ref{wca2b}) (the original
AWC expression) for the second order perturbative term in the
AWC theory, is recovered.

The correct second order AWC calculation, $[\Delta
  A]_{(2)}^\prime$ (see Eq. (\ref{wca2b})) shows that the TPT breaks
down at small reduced temperatures $\tau$ and large packing fractions
$\eta$, as expected. 

\appendix
\section{Correction to approximation (\ref{dgdrksa})}

One can understand that Eq. (\ref{dgdrksa}) is not an exact relation
by comparing the small density expansion of the left and right hand
side. For the left hand side we have \cite{Hansen}
\bq
\frac{\partial\ln g_0(1,2)}{\partial\rho}=
\mbox{\hspace{5pt}}
\put(0,0){\circle{5}}\put(20,0){\circle{5}}\put(10,17.32){\circle*{5}}
\put(0,0){\drawline(1.25,2.165)(10,17.32)}
\put(0,0){\drawline(10,17.32)(18.75,2.165)}
\put(-2.5,-15){1}\put(17.5,-15){2}
\hspace{20pt}\mbox{\hspace{5pt}}+
\left(\mbox{\hspace{5pt}}
\put(0,0){\circle{5}}\put(20,0){\circle{5}}
\put(0,20){\circle*{5}}\put(20,20){\circle*{5}}
\put(0,0){\drawline(0,2.5)(0,20)}
\put(0,0){\drawline(0,20)(20,20)}
\put(0,0){\drawline(20,20)(20,2.5)}
\put(-2.5,-15){1}\put(17.5,-15){2}
\hspace{20pt}\mbox{\hspace{5pt}} +
\mbox{\hspace{5pt}}
\put(0,0){\circle{5}}\put(20,0){\circle{5}}
\put(0,20){\circle*{5}}\put(20,20){\circle*{5}}
\put(0,0){\drawline(0,2.5)(0,20)}
\put(0,0){\drawline(0,20)(20,20)}
\put(0,0){\drawline(20,20)(20,2.5)}
\put(0,0){\drawline(0,20)(18.23,1.77)}
\put(-2.5,-15){1}\put(17.5,-15){2}
\hspace{20pt}\mbox{\hspace{5pt}} +
\mbox{\hspace{5pt}}
\put(0,0){\circle{5}}\put(20,0){\circle{5}}
\put(0,20){\circle*{5}}\put(20,20){\circle*{5}}
\put(0,0){\drawline(0,2.5)(0,20)}
\put(0,0){\drawline(0,20)(20,20)}
\put(0,0){\drawline(20,20)(20,2.5)}
\put(0,0){\drawline(20,20)(1.77,1.77)}
\put(-2.5,-15){1}\put(17.5,-15){2}
\hspace{20pt}\mbox{\hspace{5pt}} +
\mbox{\hspace{5pt}}
\put(0,0){\circle{5}}\put(20,0){\circle{5}}
\put(0,20){\circle*{5}}\put(20,20){\circle*{5}}
\put(0,0){\drawline(0,2.5)(0,20)}
\put(0,0){\drawline(0,20)(20,20)}
\put(0,0){\drawline(20,20)(20,2.5)}
\put(0,0){\drawline(20,20)(1.77,1.77)}
\put(0,0){\drawline(0,20)(18.23,1.77)}
\put(-2.5,-15){1}\put(17.5,-15){2}
\hspace{20pt}\mbox{\hspace{5pt}}\right)2\rho+O\left(\rho^2\right)~,
\eq
where in the Mayer graphs the filled circles are field points of
weight $1$ and connecting bonds are Mayer functions of the reference
system $f_0$. And using 
\bq
h_0(1,2)=
\mbox{\hspace{5pt}}
\put(0,0){\circle{5}}\put(20,0){\circle{5}}
\put(0,0){\drawline(2.5,0)(17.5,0)}
\put(-2.5,-15){1}\put(17.5,-15){2}
\hspace{20pt}\mbox{\hspace{5pt}} +
\left(\mbox{\hspace{5pt}}
\put(0,0){\circle{5}}\put(20,0){\circle{5}}\put(10,17.32){\circle*{5}}
\put(0,0){\drawline(1.25,2.165)(10,17.32)}
\put(0,0){\drawline(10,17.32)(18.75,2.165)}
\put(0,0){\drawline(2.5,0)(17.5,0)}
\put(-2.5,-15){1}\put(17.5,-15){2}
\hspace{20pt}\mbox{\hspace{5pt}} +
\mbox{\hspace{5pt}}
\put(0,0){\circle{5}}\put(20,0){\circle{5}}\put(10,17.32){\circle*{5}}
\put(0,0){\drawline(1.25,2.165)(10,17.32)}
\put(0,0){\drawline(10,17.32)(18.75,2.165)}
\put(-2.5,-15){1}\put(17.5,-15){2}
\hspace{20pt}\mbox{\hspace{5pt}}\right)\rho+O\left(\rho^2\right)~,
\eq
in the right hand side one finds,
\bq
&&\frac{\chi_T^{id}}{\chi_T^0}\int d3\,h_0(1,3)h_0(2,3)
=\frac{\int d3\,h_0(1,3)h_0(2,3)}{1+\displaystyle\frac{\rho}{V}\int
d1d2\,h_0(1,2)}\\[1cm] \nonumber
&&=\frac{
\mbox{\hspace{5pt}}
\put(0,0){\circle{5}}\put(20,0){\circle{5}}\put(10,17.32){\circle*{5}}
\put(0,0){\drawline(1.25,2.165)(10,17.32)}
\put(0,0){\drawline(10,17.32)(18.75,2.165)}
\put(-2.5,-15){1}\put(17.5,-15){2}
\hspace{20pt}\mbox{\hspace{5pt}}+
\left(\mbox{\hspace{5pt}}
\put(0,0){\circle{5}}\put(20,0){\circle{5}}
\put(0,20){\circle*{5}}\put(20,20){\circle*{5}}
\put(0,0){\drawline(0,2.5)(0,20)}
\put(0,0){\drawline(0,20)(20,20)}
\put(0,0){\drawline(20,20)(20,2.5)}
\put(-2.5,-15){1}\put(17.5,-15){2}
\hspace{20pt}\mbox{\hspace{5pt}} +
\mbox{\hspace{5pt}}
\put(0,0){\circle{5}}\put(20,0){\circle{5}}
\put(0,20){\circle*{5}}\put(20,20){\circle*{5}}
\put(0,0){\drawline(0,2.5)(0,20)}
\put(0,0){\drawline(0,20)(20,20)}
\put(0,0){\drawline(20,20)(20,2.5)}
\put(0,0){\drawline(0,20)(18.23,1.77)}
\put(-2.5,-15){1}\put(17.5,-15){2}
\hspace{20pt}\mbox{\hspace{5pt}} +
\mbox{\hspace{5pt}}
\put(0,0){\circle{5}}\put(20,0){\circle{5}}
\put(0,20){\circle*{5}}\put(20,20){\circle*{5}}
\put(0,0){\drawline(0,2.5)(0,20)}
\put(0,0){\drawline(0,20)(20,20)}
\put(0,0){\drawline(20,20)(20,2.5)}
\put(0,0){\drawline(20,20)(1.77,1.77)}
\put(-2.5,-15){1}\put(17.5,-15){2}
\hspace{20pt}\mbox{\hspace{5pt}}\right)2\rho+O\left(\rho^2\right)
}{\displaystyle
1+
\left(
\frac{\mbox{\hspace{5pt}}
\put(0,0){\circle*{5}}\put(20,0){\circle*{5}}
\put(0,0){\drawline(2.5,0)(17.5,0)}
\hspace{20pt}\mbox{\hspace{5pt}}
}{V}\right)2\rho
+O\left(\rho^2\right)
}\\[1cm] \nonumber
&&=
\mbox{\hspace{5pt}}
\put(0,0){\circle{5}}\put(20,0){\circle{5}}\put(10,17.32){\circle*{5}}
\put(0,0){\drawline(1.25,2.165)(10,17.32)}
\put(0,0){\drawline(10,17.32)(18.75,2.165)}
\put(-2.5,-15){1}\put(17.5,-15){2}
\hspace{20pt}\mbox{\hspace{5pt}}+
\left[\mbox{\hspace{5pt}}
\put(0,0){\circle{5}}\put(20,0){\circle{5}}
\put(0,20){\circle*{5}}\put(20,20){\circle*{5}}
\put(0,0){\drawline(0,2.5)(0,20)}
\put(0,0){\drawline(0,20)(20,20)}
\put(0,0){\drawline(20,20)(20,2.5)}
\put(-2.5,-15){1}\put(17.5,-15){2}
\hspace{20pt}\mbox{\hspace{5pt}} +
\mbox{\hspace{5pt}}
\put(0,0){\circle{5}}\put(20,0){\circle{5}}
\put(0,20){\circle*{5}}\put(20,20){\circle*{5}}
\put(0,0){\drawline(0,2.5)(0,20)}
\put(0,0){\drawline(0,20)(20,20)}
\put(0,0){\drawline(20,20)(20,2.5)}
\put(0,0){\drawline(0,20)(18.23,1.77)}
\put(-2.5,-15){1}\put(17.5,-15){2}
\hspace{20pt}\mbox{\hspace{5pt}} +
\mbox{\hspace{5pt}}
\put(0,0){\circle{5}}\put(20,0){\circle{5}}
\put(0,20){\circle*{5}}\put(20,20){\circle*{5}}
\put(0,0){\drawline(0,2.5)(0,20)}
\put(0,0){\drawline(0,20)(20,20)}
\put(0,0){\drawline(20,20)(20,2.5)}
\put(0,0){\drawline(20,20)(1.77,1.77)}
\put(-2.5,-15){1}\put(17.5,-15){2}
\hspace{20pt}\mbox{\hspace{5pt}} -
\left(\mbox{\hspace{5pt}}
\put(0,0){\circle{5}}\put(20,0){\circle{5}}\put(10,17.32){\circle*{5}}
\put(0,0){\drawline(1.25,2.165)(10,17.32)}
\put(0,0){\drawline(10,17.32)(18.75,2.165)}
\put(-2.5,-15){1}\put(17.5,-15){2}
\hspace{20pt}\mbox{\hspace{5pt}}\right)\cdot
\left(\frac{\mbox{\hspace{5pt}}
\put(0,0){\circle*{5}}\put(20,0){\circle*{5}}
\put(0,0){\drawline(2.5,0)(17.5,0)}
\hspace{20pt}\mbox{\hspace{5pt}}}{V}\right)
\right]2\rho + O\left(\rho^2\right)\\[1cm] \nonumber
&&=
\alpha_0(1,2) + \alpha_1(1,2) + O\left(\rho^2\right)~,
\eq
So that the correction term is of order $\rho$, namely,
\bq
\alpha_1^\prime(1,2)=
\left[\mbox{\hspace{5pt}}
\put(0,0){\circle{5}}\put(20,0){\circle{5}}
\put(0,20){\circle*{5}}\put(20,20){\circle*{5}}
\put(0,0){\drawline(0,2.5)(0,20)}
\put(0,0){\drawline(0,20)(20,20)}
\put(0,0){\drawline(20,20)(20,2.5)}
\put(0,0){\drawline(20,20)(1.77,1.77)}
\put(0,0){\drawline(0,20)(18.23,1.77)}
\put(-2.5,-15){1}\put(17.5,-15){2}
\hspace{20pt}\mbox{\hspace{5pt}} +
\left(\mbox{\hspace{5pt}}
\put(0,0){\circle{5}}\put(20,0){\circle{5}}\put(10,17.32){\circle*{5}}
\put(0,0){\drawline(1.25,2.165)(10,17.32)}
\put(0,0){\drawline(10,17.32)(18.75,2.165)}
\put(-2.5,-15){1}\put(17.5,-15){2}
\hspace{20pt}\mbox{\hspace{5pt}}\right)\cdot
\left(\frac{\mbox{\hspace{5pt}}
\put(0,0){\circle*{5}}\put(20,0){\circle*{5}}
\put(0,0){\drawline(2.5,0)(17.5,0)}
\hspace{20pt}\mbox{\hspace{5pt}}}{V}\right)
\right]2\rho~.
\eq
The correct small density expansion for the density derivative of the
two body correlation function is 
\bq
\frac{\partial\rho_0(1,2)}{\partial\rho}=
g_0(1,2)\left[2\rho+\rho^2\alpha_0(1,2)+\rho^2\alpha_1(1,2)+
\rho^2\alpha_1^\prime(1,2)+O\left(\rho^4\right)\right]~,
\eq
where the first term neglected in KSA is $\rho^2\alpha_1^\prime=O(\rho^3)$.  

%


\end{document}